\documentclass{article}
\usepackage{epsfig}

\usepackage{amsmath,amssymb}
\usepackage[latin1]{inputenc}
\usepackage{colortbl}
\usepackage[english]{babel}
\usepackage[dvips]{rotating}
\usepackage{epsf}
\usepackage{psfrag}

\usepackage{times}

\newcommand{\rf}[1]{(\ref{#1})}
\newcommand{\beq}{\begin{equation}}

\newcommand{\eeq}{\end{equation}}
\newcommand{\bea}{\begin{eqnarray}}
\newcommand{\eea}{\end{eqnarray}}

\newcommand{\e}{\mbox{e}}
\renewcommand{\d}{\mbox{d}}

\newcommand{\lam}{\lambda}
\newcommand{\Lam}{\Lambda}
\renewcommand{\b}{\beta}
\renewcommand{\a}{\alpha}

\newcommand{\ep}{\varepsilon}     

\newcommand{\om}{\omega}

\newcommand{\del}{\delta}

\newcommand{\oh}{\frac{1}{2}}

\newcommand{\dg}{\dagger}

\newcommand{\tr}{\mathrm{tr}\,}
\newcommand{\Tr}{\mathrm{Tr}\,}
\newcommand{\ra}{\rangle}
\newcommand{\la}{\langle}
\newcommand{\lan}{\left\la}
\newcommand{\ran}{\right\ra}

\newcommand{\prt}{\partial}
\newcommand{\mi}{\!-\!}

\newcommand{\plu}{\!+\!}

\newcommand{\tg}{{\tilde{g}}}

\newcommand{\hI}{{\hat{I}}}

\newcommand{\bg}{{\bar{g}}}
\newcommand{\bz}{{\bar{z}}}
\newcommand{\bV}{{\bar{V}}}

\newcommand{\cdtL}{\Lam_{{\rm cdt}}}
\newcommand{\cdtZ}{Z_{\rm cdt}}
\newcommand{\cdtW}{W_{\rm cdt}}

\begin{document}

\begin{center}
\vspace{24pt}
{ \large \bf CDT as a scaling limit of matrix models}\footnote{Based 
on a talk presented at the XXIII Marian Smoluchowski Symposium 
on Statistical Physics,
``Random Matrices, Statistical Physics and Information Theory'', 
Krakow, Poland, September 26-30, 2010. To appear in 
Acta Physica  Polonica B, Vol 42, No 5 (2011) .
}

\vspace{30pt}

{\sl J. Ambj\o rn}

\vspace{24pt}
{\footnotesize

The Niels Bohr Institute, Copenhagen University\\
Blegdamsvej 17, DK-2100 Copenhagen \O , Denmark.\\
{ email: ambjorn@nbi.dk}\\

}
\vspace{48pt}

\end{center}


\begin{center}
{\bf Abstract}
\end{center}

It is shown that generalized CDT, 
the two-dimensional theory of quantum gravity,
constructed as a scaling limit from so-called causal dynamical triangulations,
can be obtained from a cubic matrix model. It involves taking a
new scaling limit of matrix models, which  is  more natural 
from a classical point of view.


\section*{Introduction}\label{sec1}

The great versatility of matrix models or matrix integrals in theoretical
physics is well illustrated by their particularly 
beautiful application in two-dimensional 
Euclidean quantum gravity (see \cite{david1,gm,difgz,book}  
for reviews). This theory 
can be defined as a suitable sum over triangulations, so-called 
``dynamical triangulations'' (DT), whose continuum
limit is obtained by taking the side lengths $a$ 
of the triangles to zero. 
The method of DT was originally introduced as a nonperturbative worldsheet 
regularization of the Polyakov bosonic string \cite{ambjorn,david,mkk}.
There it was used with success 
(or to disappointment, depending on ones taste) 
to show rigorously that a tachyon-free version
of Polyakov's bosonic string theory does not exist in target 
space dimensions $d > 1$ \cite{ad}. 
However, when viewed as a theory of 2d quantum 
gravity coupled to matter with central charge $c\leq 1$, 
the theory -- noncritical string theory -- is perfectly consistent,
and matrix models have been used to solve
in an elegant way the combinatorial aspects of the 
DT construction, where one sums over random surfaces
glued together from equilateral triangles.

The DT approach possesses a well-defined cut-off, the 
length $a$ of the lattice links. As has been discussed in many
reviews (for instance the ones mentioned above), 
a continuum limit can be defined when the 
lattice spacing is taken to zero while simultaneously renormalizing
the bare cosmological constant and possibly other matter
coupling constants. However, the continuum limit in question has
some unconventional properties. We will
show that there is another way of taking the scaling limit of 
the matrix models, which still relates them to a summation
over triangulated random surfaces, the so-called 
{\it causal dynamical triangulations} (CDT) \cite{al}. We will
show that this limit is in a way more natural and 
it corresponds to starting out from a ``classical'' 
matrix-model theory. In accordance with this it 
does not lead to the somewhat unconventional renormalization
encountered in the standard DT approach.

\section*{Hermitian Matrix Models}

We can define the Hermitian matrix model for $N\times N$ matrices 
as a formal power series in $\tilde{g}$
\bea
Z(\tilde{g}) &=& \int d\phi \; 
\e^{-N \tr \left(\oh \phi^2 -\frac{\tg}{3} \phi^3\right)}\label{1.1} \\
&=& \sum_{k=0}^\infty \frac{1}{ k!}\;
\int d\phi \; \e^{-\oh N \tr \left( \phi^2\right)}\; 
 \left( \frac{N\tg}{3} \tr \phi^3\right)^k, 
\nonumber
\eea
\beq\label{1.2}
d\phi =\prod_{\a \leq \b} d\, {\rm Re}\, \phi_{\a\b}
\prod_{\a < \b} d \,{\rm Im}\, \phi_{\a\b}.
\eeq

The integral can be evaluated in the standard way
by performing  all possible Wick contractions of $(\Tr\phi^3)^k$ and using
\beq\label{1.3}
\lan \phi_{\a\b} \phi_{\a'\b'} \ran =C
\int d \phi \; e^{-\oh \sum_{\a\b}|\phi_{\a\b}|^2} \phi_{\a\b}
\phi_{\a'\b'}
= \del_{\a\b'} \del_{\b\a'},
\eeq
\begin{figure}[t]
\centerline{\scalebox{0.4}{\rotatebox{0}{\includegraphics{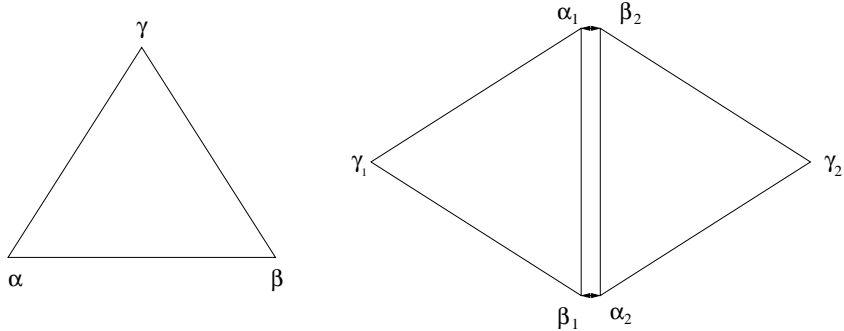}}}}
\caption{The gluing of triangles via Gaussian integration.}
\label{fig1}
\end{figure}

The geometric interpretation in the context of DT is illustrated in 
Fig.\ \ref{fig1}. An index is assigned with each vertex
in a triangle and a matrix $\phi_{\a\b}$ is assigned to the link
(the side in the triangle) which contains the vertices labeled $\a$ and $\b$.
In that way we can associate $\tr \phi^3$ with a triangle and the term 
$(\phi^3)^k$  in \rf{1.1} with $k$ triangles. Performing the 
Gaussian integrations indicated in eq.\ \rf{1.3} we are gluing 
together all triangles in all possible ways by identifying  links 
as illustrated in the figure. This way of performing the integral 
will result in an asymptotic power series in $\tg$ which can be Borel 
summed for $\tg < 0$.

If we want to perform the integral without power expanding the $\phi^3$
part of the potential one takes advantage of the invariance of the
action under $\phi \to U \phi U^\dg$, where $U \in U(N)$. Thus the action 
depends only on the eigenvalues $\ell_i$ of $\phi$ and we can make 
the following decomposition of the measure of integration:
\beq\label{1.4}
d \phi \;\e^{-\frac{N}{g_s} \tr V(\phi)} 
\propto \d U(N) \prod_{i=1}^N d \ell_i \; \e^{-\frac{N}{g_s} V(\ell_i)}
\prod_{i<j} |\ell_i-\ell_j|^2,
\eeq
where $\prod_{i<j} |\ell_i-\ell_j|^2$ is the Jacobian, changing from
$\phi$ to its eigenvalues and the unitary matrix $U$. The integral over 
the $U$ matrices is now trivial since the action is independent of $U$.  

The ``{classical}'' limit is obtained for $g_s \to 0$,
where all eigenvalues are lumped together at $\ell_0$, where 
\beq\label{1.5}
V'(\ell_0)=0.
\eeq

However, for $g_s > 0$ the integration over the non-diagonal matrix 
elements produces the Vandermonde determinant $\prod_{i<j} |\ell_i-\ell_j|^2$, 
which acts as a ``quantum'' correction, a repulsion between different 
eigenvalues. The result is that eigenvalues are no longer
concentrated at $\ell_0$. In the large $N$ limit they occupy an
interval around $\ell_0$.

\begin{figure}[t]
\centerline{\scalebox{0.4}{\rotatebox{0}{\includegraphics{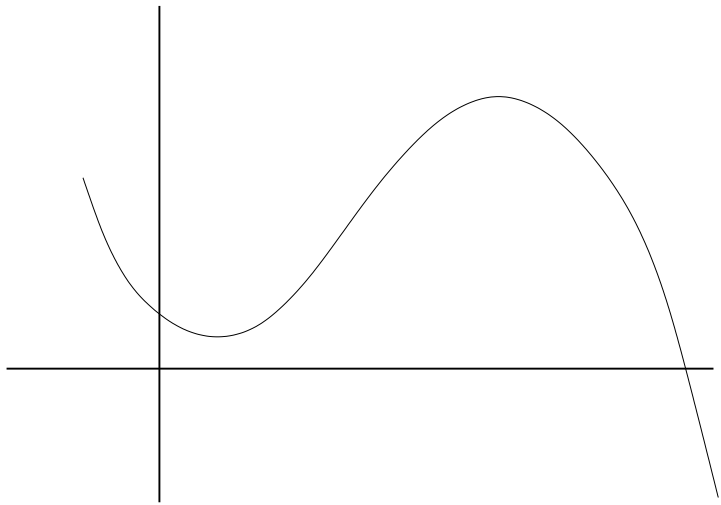}}}}
\caption{The graph $-g\phi + \oh \phi^2 - \frac{g}{3} \phi^3$}
\label{fig2}
\end{figure}

As an example we let us consider the matrix potential 
\beq\label{1.6}
\frac{1}{g_s}V(\phi) = \frac{1}{g_s}
\Big(-g\phi + \oh \phi^2 - \frac{g}{3} \phi^3\Big) 
\eeq
shown in Fig.\ \ref{fig2}. It is clear that this specific  matrix integral 
only exists as the formal power series for finite $N$. For infinite $N$
the eigenvalues will condense in an interval around $\ell_0$. 
This interval is determined by a large $N$ saddelpoint equation.
We will not here discuss the solution to that equation, but only 
mention that the so-called resolvent, which determines the distribution
of eigenvalues is given by:
\beq\label{1.7}
w(z):= \lan \frac{1}{N} \tr \frac{1}{z-\phi} \ran = \frac{1}{Z} 
\int d \phi \; \frac{1}{N} \frac{1}{z-\phi} \; \e^{-\frac{N}{g_s}\tr V(\phi)} 
\eeq

For $V(\phi) = -g\phi + \oh \phi-\frac{g}{3} \phi^3$ one has from the 
saddelpoint equation or by other methods, like the so-called loop equations, 
to leading order in $N$:
\beq\label{1.8}
w(z) = \frac{1}{2g_s} \left(V'(z) + g(z-b)\sqrt{(z-c)(z-d)}\right),
\eeq
where the constants $b$, $c$ and $d$ are determined by the requirement 
that $w(z) \to 1/z$ for $z \to \infty$. It should further be noticed that
in the large $N$ expansion each term has analyticity structure like 
$w(z)$ in the complex $z$-plane, i.e.\ a branch cut between $c$ and $d$.

\section*{The conventional scaling limit}

The usual scaling limit of the matrix model, relevant for 
non-critical strings and 2d Euclidean quantum gravity coupled 
to matter,  is obtained (for fixed $g_s$) by fine-tuning 
$g$ such that $b(g)=c(g)$. At this point the analytic 
structure of $w(z)$ changes from  
$(z-c(g))^{1/2}\to (z-c(g))^{3/2}$,
and this change can only be accommodated by invoking arbitrary 
high $k$ in the sum   
\beq\label{2.1}
\sum_{k=0}^\infty \frac{1}{ k!}\;
\int d\phi \; \e^{-\frac{N}{2g_s} \tr \left( \phi^2\right)}\; 
 \left( \frac{N g}{3g_s} \tr \phi^3\right)^k, 
\eeq
This is why one geometrically can imagine a ``continuum'' limit 
where the size of each triangle shrinks to zero while the continuum
size of the surface stays constant. More precisely, for our specific 
model one has:
\beq\label{2.2}
g = g_c(1-\Lam a^2),~~~~~z = c(g_c)+a Z,~~~~~a\to 0,
\eeq
\beq\label{2.3}
w(z) = \oh \left(V'(z) + g \sqrt{c(g_c)-d(g_c)}\; a^{3/2} 
W_E(Z,\Lam)+ 0(a^{5/2})\right),
\eeq
where the ``continuum disk amplitude is 
\beq\label{2.4}
W_E(Z,\Lam) = (Z- \sqrt{2\Lam/3}) \sqrt{Z+2\sqrt{2\Lam/3}}.
\eeq
Here $a$ has the interpretation as the length of the side 
of the triangles (polygons) which appear in $V(\phi)$.

Notice that actually the non-scaling part
$V'(z)/2$ dominates when $a \to 0$ and  renders
the average number of polygons present in the ensemble with partition
function $w(z)$  finite, even at the critical point. 
This somewhat embarrassing fact can be circumvented by differentiating 
$w(z)$ a sufficient number of times with respect to  
$g$ and $z$, after which these ``non-universal'' 
contributions vanish since they are polynomials in $g$ and $z$, but for 
the disk amplitude itself there is no such escape.

\section*{The new scaling limit}

Until now we have considered two limits: 
the ``classical limit'': $g_s =0$ and ``conventional scaling  
limit'' of non-critical string theory: $g_s > 0$ and $g\to g_c (g_s)$.
Is it possible to find a new, non-trivial scaling limit, closer
to the classical limit when $g_s \to 0$. The answer is yes \cite{newscaling}.

If one works out the details close to the conventional
critical point $b(g_c)=c(g_c)$ one has
\beq\label{3.1}
g_c(g_s)= \oh (1- 
\frac{3}{2} g_s^{2/3} + O(g_s^{4/3})), 
\eeq
\beq\label{3.2}
z_c(g_s)=c(g_c,g_s)= 
1 + g_s^{1/3} + O(g_s^{2/3}),
\eeq
\beq\label{3.3}
c(g_c)-d(g_c) = 4 g_s^{1/3} + 0(g^{2/3}_s)
\eeq

A non-trivial scaling can now be obtained if we 
fine-tune $g_s \to 0$ as 
\beq\label{gs}
g_s = G_s a^3.
\eeq
Again the scaling parameter $a$ can be given the geometric 
interpretation as the link lengths of the polygons in $V(\phi)$.
Note that the length of the cut goes to zero as $a \to 0$, thus we 
are closer to the ``classical'' limit. However, it will survive in the 
continuum limit:
\beq\label{3.4}
g = g_c(g_s)(1-a^2 \Lam) = \bar{g}(1 - a^2 \cdtL+O(a^4))
\eeq
\beq\label{3.5}
z= z_c +a Z = \bz+a \cdtZ +O(a^2)
\eeq
\beq\label{3.6}
\cdtL \equiv \Lam + \frac{3}{2}G_s^{2/3},~~~\bg = \oh,~~~
\cdtZ\equiv Z+G_s^{1/3},~~~\bz=1.
\eeq
Using these definitions one computes in the limit $a \to 0$ that
\beq\label{3.7}
w(z) = \frac{1}{a} \; \frac{\cdtL -\oh \cdtZ^2 + 
\oh(\cdtZ-H)\sqrt{(\cdtZ+H)^2 -\frac{4G_s}{H}}}{2G_s}.
\eeq
\beq\label{3.8}
h^3 -h + \frac{2G_s}{(2\cdtL)^{3/2}} =0,~~~h=H/\sqrt{2\cdtL}
\eeq
\beq\label{3.9}
w(z) = \frac{1}{a}\, \cdtW(\cdtZ,\cdtL,G_s) 
\eeq
Thus we have a situation where, contrary the situation we encountered 
taking the conventional scaling limit discussed above, the disk amplitude
$w(z)$ really scales as one would expect from an ordinary correlation 
function. 

We can now take the limit $G_s \to 0$ and we obtain
\beq\label{3.10}
\cdtW(\cdtZ,\cdtL,G_s) \to \frac{1}{\cdtZ +\sqrt{2\cdtL}}.
\eeq
Thus we see that the cut where the eigenvalues are located shrinks to 
a point, indicating we have precisely the classical situation discussed 
above. This is indeed the case as we will discuss further shortly.

If we on the other hand take the limit $G_s \to \infty$ we obtain
for the square root part of $\cdtW(\cdtZ,\cdtL,G_s)$
\beq\label{3.11}
\frac{(\cdtZ\mi H)\sqrt{(\cdtZ\plu H)^2 \mi\frac{4G_s}{H}}}{2G_s} \to
G_s^{-5/6}\; 
\left(Z\mi \sqrt{2\Lam/3}\right)\sqrt{Z\plu 2\sqrt{2\Lam/3}}\nonumber
\eeq
Thus we recover the standard continuum disk function $W_E(Z,\Lam)$
from \rf{2.4} times a 
factor $G_s^{-5/6}$. If we write $g_s = a^3G_s$ (in accordance with
\rf{gs}) and keep $g_s$ 
constant, while taking $a \to 0$, it means that $G_s \to \infty$ as 
$a^{-3}$. Thus $G_s^{-5/6} \sim a^{5/2}$ and we precisely 
recover the square root term in \rf{2.3} if we remember
that $w(z)$ and $\cdtW $ according to \rf{3.9} differ by a
factor $a$.
However, the part not related to the square root in 
$\cdtW$ will not scale in the limit $G_s \to \infty$, in accordance
with the previous discussion of standard scaling related to \rf{2.3}.  

\section*{The matrix model}

The new scaling can be obtained by a simple change of variables:
\beq\label{4.1}
\phi \to \bz \, \hI + a \Phi + O(a^2)
\eeq
Up to a $\phi$ independent term we then have
\beq\label{4.2}
V(\phi) = \bV(\Phi),~~~~
\bV(\Phi) \equiv \frac{\cdtL \Phi -\frac{1}{6} \Phi^3}{2G_s},
\eeq
where $V(\phi)$ is the potential given by \rf{1.6}. 
Thus we can write
\beq\label{4.3}
Z(g,g_s) = a^{N^2} Z(\cdtL,G_s),~~~Z(\cdtL,G_s):=
\int d\Phi \; 
\e^{-N\tr \bV(\Phi)}
\eeq
\begin{figure}[t]
\centerline{\scalebox{0.4}{\rotatebox{0}{\includegraphics{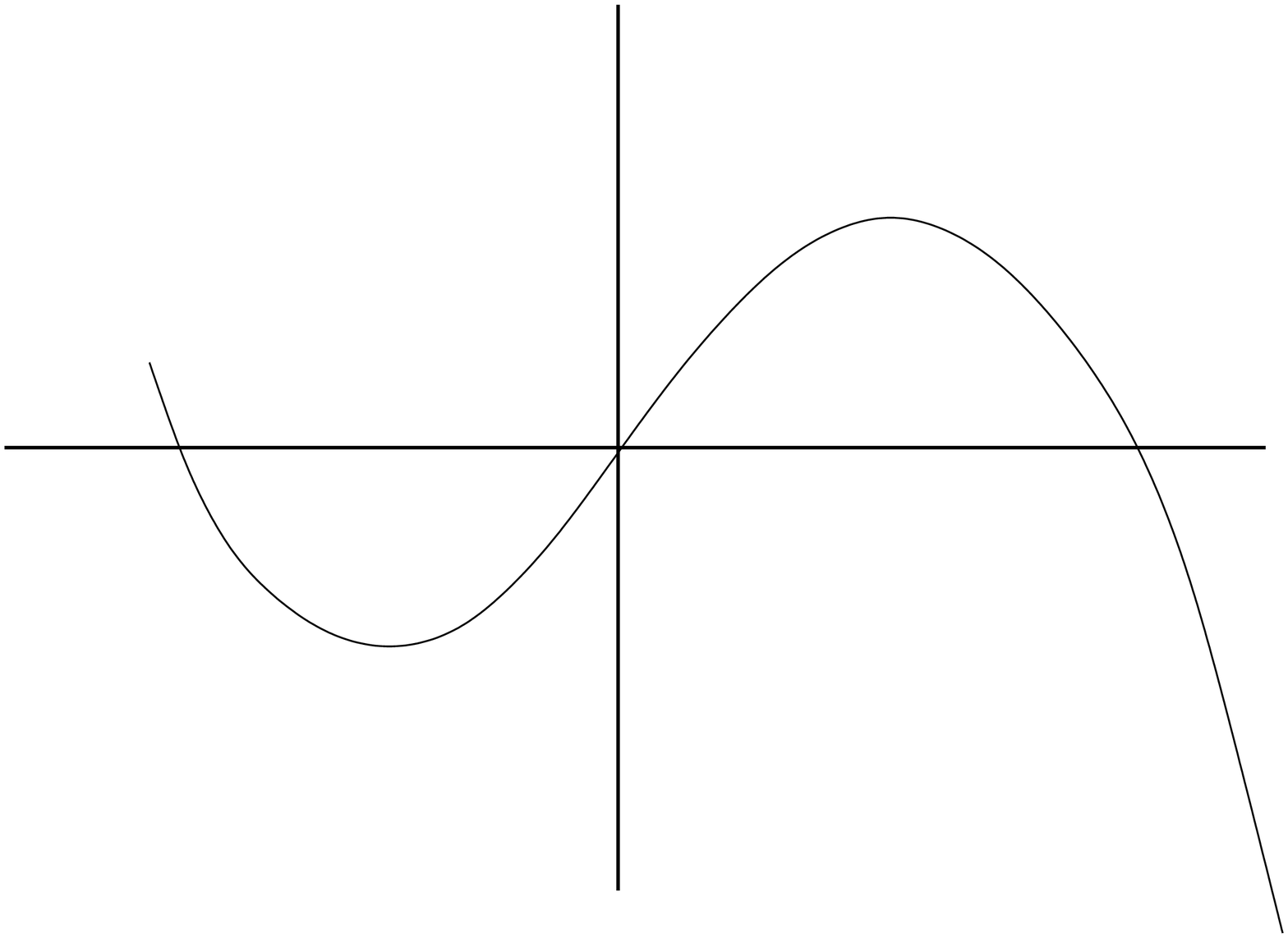}}}}
\caption{The potential $\bV(\ell)$, \rf{4.2}, with the local minimum 
$\ell_0= -\sqrt{2\cdtL}$.}
\label{fig3}
\end{figure}
The change of variable \rf{4.1} explains immediately and in a trivial 
way the scaling \rf{3.9} of $w(z)$:
\beq\label{4.4}
\frac{1}{z-\phi}= \frac{1}{a}\, \frac{1}{\cdtZ-\Phi} ~~~~\Rightarrow~~~
w(z)=\frac{1}{a}  \cdtW(\cdtZ,\cdtL,G_s)
\eeq
This relation is obviously correct to all orders in $N$
What is truly surprising is that  the new scaling limit
is itself a matrix model defined by $\bV(\Phi)$ \cite{contmatrix}. 
The continuum limit $a\to 0$ is thus described by a matrix model. This 
bear some resemblance with  the Kontsevich matrix model,
even the matrix potential is somewhat similar.
But contrary to that model, where the continuum objects 
are the modular spaces of Riemann surfaces it turns out 
that the present matrix model actually describes a set 
of ``real'' continuum surfaces as we will describe below. 

We have  
\beq\label{4.5}
\bV(\Phi) \propto 2\cdtL \Phi - \frac{1}{3}\Phi^3,
\eeq
and the cubic potential is shown in Fig.\ \ref{fig3}. It has a local minimum
$\ell_0$ determined by  
\beq\label{4.6}
\bV'(\ell_0)=0~~~\Rightarrow ~~~ \ell_0= -\sqrt{2\cdtL}.
\eeq

Thus the ``classical'' limit of the matrix integral with potential 
$\bV(\Phi)$, when only the minimum plays a role, leads 
to the following expectation value:
\beq\label{4.7}
 \frac{1}{N} \lan  \tr \frac{1}{\cdtZ-\Phi} \ran = 
\frac{1}{\cdtZ+\sqrt{2\cdtL}}
=\lim_{G_s \to 0} \cdtW(\cdtZ,\cdtL,G_s), 
\eeq
so in this way one can use this matrix model very explicitly 
to obtain the classical limit. As we will show in the next Section
even this classical limit has a non-trivial representation as 
a sum over certain random surfaces.

\section*{Geometric interpretation}

Let us define some geometric objects, related by Laplace transformations,
$W_{\lam,g_s}(\ell)$ and $W_{\lam,g_s}(x)$. We have 
now a little abuse of notation. Above we used capital letters
for continuum, dimensionful variables and small letters for dimensionless
(lattice) variables. Now we will use $\lam$ and $g_s$ $x$ for continuum,
dimensionful variables. In the end they will be identified with
the continuum variables $\cdtL$, $G_s$ and $\cdtZ$ etc., in the same 
way as the object  $W_{\lam,g_s}(x)$ in the end will be identified
with $\cdtW (\cdtZ,\cdtL, G_s)$. However, we do 
it in order to stress that we are now starting from scratch with 
a geometric theory which in principle has nothing to do with the matrix 
model above. And in fact that was how these ``quantum geometric'' objects 
to  first found and analyzed \cite{cdt1}-\cite{cdt2}.  
\beq\label{5.1}
W_{\lam,g_s} (x) = \int_0^\infty d \ell \; \e^{-x \ell}\; W_{\lam,g_s}(\ell)
\eeq
\begin{figure}[t]
\centerline{\scalebox{0.5}{\rotatebox{0}{\includegraphics{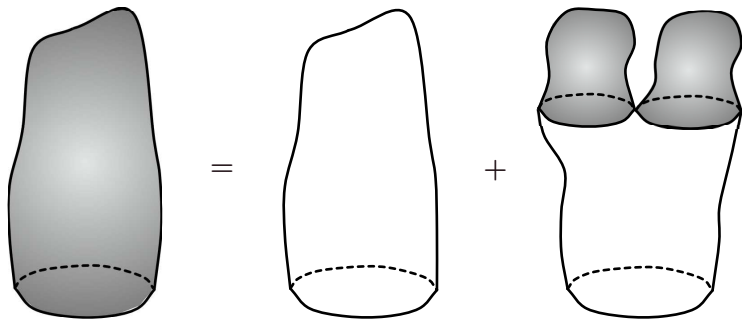}}}}
\caption{The graphic representation of the integral equation 
satisfied by the full disk amplitude, eq.\ \rf{5.2}.}
\label{fig4}
\end{figure}
The objects are intended to represent the disk amplitude in the 
theory of two-dimensional quantum gravity based on causal dynamical 
triangulated random surfaces (CDT). The idea is to start be summing 
over all random surfaces with a boundary of length $\ell$ which admit 
a proper time foliation. The action used is just the 
area action (i.e.\ the cosmological term, since the 
Einstein term is trivial in 2d as long as we do not allow
topology change. And we will not allow that presently. 
Thus the topology of the surface is just the trivial topology of the 
disk.) We denote this sum the disk amplitude $W_\lam^{(0)}(\ell)$.
It is called the CDT disk amplitude. If we add a boundary cosmological 
constant $x$ at the boundary, we should add a boundary action $x\, \ell$,
and thus the Laplace transformation \rf{5.1} can be viewed as changing 
the path integral over surfaces with a fixed boundary length to a path
integral where we also integrate over all boundary lengths and instead 
keep fixed a boundary cosmological constant $x$.

We now allow a larger class of surfaces by allowing branching, i.e.\ we 
allow the spatial surface at a proper time $t$ to split in two, the 
splitting assigned a weight $g_s$. $g_s$ is clearly like a string 
coupling constant. The process is shown in Fig.\ \ref{fig4}. 
We are allowing the splitting of a spatial universe in two, but 
presently we do not allow for topology change of the two-dimensional 
surface, so we do not allow the two universes to join again, since 
that would create a handle and then change the two-dimensional topology.
Had we allowed it, the total coupling constant associated with this 
process would have been $g_s^2$, one factor for splitting, one factor
for joining, like in string theory.
The unshaded disk amplitude is  $W_\lam^{(0)}(\ell)$, while the 
full disk amplitude is denoted  $W_{\lam,g_s}(\ell)$ and is shown
as the shaded graph on the lhs of the equality sign. 

Fig.\ \ref{fig4} is a graphic representation of the following 
integral equation:
\bea
\lefteqn{W_{\lam,g_s} (x) = W_{\lam} ^{(0)}(x) +} \label{5.2}\\
&& g_s\int\limits_0^\infty dt \int\limits_0^\infty d \ell_1 d \ell_2  \;
(\ell_1+\ell_2) G^{(0)}_{\lam}  (x,\ell_1+\ell_2;t) W_{\lam,g_s} (\ell_1)
W_{\lam,g_s} (\ell_2)
\nonumber 
\eea
In this equation the object $G^{(0)}_{\lam}  (\ell_1,\ell_2;t)$ denotes 
the ``propagator'' in CDT. It is defined in analogy with 
$W_{\lam}^{(0)}(\ell)$. We sum over all two-dimensional geometries 
where we have a spatial entrance loop of length $\ell_1$ and a spatial 
exit loop of length $\ell_2$, with the further constraint that all point 
on the exit loop is separated a geodesic distance $t$ from the entrance loop.
Again we assume that all points separated a geodesic
distance $t' \leq t$ from the entrance loop form a connected one-dimensional
space. $G^{(0)}_{\lam}  (x,\ell';t)$ denotes the Laplace transform
of $G^{(0)}_{\lam}  (\ell,\ell';t)$ with respect to $\ell$.
\begin{figure}[t]
\centerline{\scalebox{0.4}{\rotatebox{0}{\includegraphics{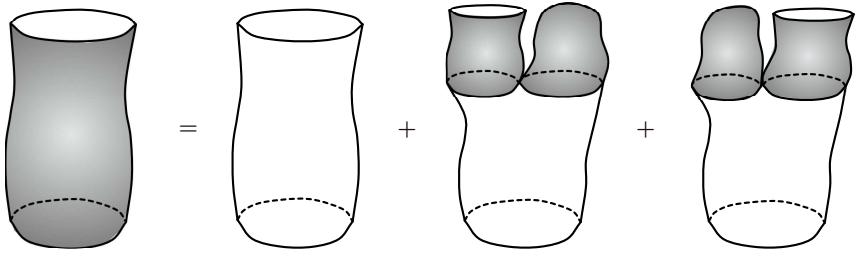}}}}
\caption{The graphic illustration of the geometries which contribute
to the propagator $G_{\lam,g_s}(x,y;t)$ in eq.\ \rf{5.4}. }
\label{fig5}
\end{figure}

In the same way as we generalized the geometries which
entered into path integral defining $W_\lam^{(0)}(\ell)$, by 
allowing space to split, and in this was obtained $W_{\lam,g_s}(\ell)$,
we can allow for the spatial hyper-surface at any time $t' \leq t$ 
to separate in two. One of these will then be connected to the exit loop
while the other will eventually vanish in the vacuum. This is shown
in Fig.\ \ref{fig5}. We denote this generalized propagator 
$G_{\lam,g_s}(\ell_1,\ell_2;t)$. For both  $G_{\lam,g_s}(\ell_1,\ell_2;t)$
and $G_{\lam}^{(0)}(\ell_1,\ell_2;t)$ we introduce, again in analogy
with the definitions for $W_{\lam,g_s}(\ell)$ and $W_{\lam}^{(0)}(\ell)$,
the Laplace transforms:
\beq\label{5.3}
G_{\lam,g_s}(x,y;t) = \int_0^\infty d \ell_1 \int_0^\infty d \ell_2\; 
\e^{-x \ell_1}\e^{\ell_2y\;}G_{\lam,g_s}(\ell_1,\ell_2;t),
\eeq
as well as the hybrid forms $G_{\lam,g_s}(x,\ell_2;t)$ and 
$G_{\lam,g_s}(\ell_1,y;t)$. Here $x$ and $y$ denotes boundary
cosmological constants at the entry and exit boundaries. 

The shaded parts of graphs in Fig.\ \ref{fig5} represent
the full, $g_s$-dependent propagator and the full $g_s$-dependent 
disc amplitude, and the non-shaded 
parts the CDT propagator where $g_s = 0$.
In all four graphs, the geodesic distance from the final to the initial 
loop is given by $t$, while the ``baby-universes'' which end in the vacuum
can terminate their life at any positive time after they have been created,
also at a time {\it larger} than $t$.

From Fig.\ \ref{fig5} one can write
down an integral equation much like eq.\ \rf{5.2} for $W_{\lam,g_s}(x)$.
However it is convenient the differentiate this equation with respect to 
$t$ and we then obtain: 
\beq\label{5.4}
 a^{\ep}\frac{\prt}{\prt t} G_{\lam,g_s}(x,y;t) = 
- \frac{\prt}{\prt x} \Big[\Big(a(x^2-\lam)+2 g_s a^{\del}\, a^{\eta-1} 
W_{\lam,g_s}(x)\Big) 
G_{\lam,g_s}(x,y;t)\Big],
\eeq
In this equation we have explicitly assumed that we 
have some kind of regularized theory, on a lattice, say. 
The equation is then first written in terms of the dimensionless 
lattice variables and the translated into continuum notation
by inserting the relation between the dimensionless variables and
their continuum counterparts, in this way introducing the lattice 
cut-off $a$. This cut-off will then appear with a power determined
by the dimension of the continuum variables, except for a subtlety related
to $W$, to be explained now. The lattice cut-off $a$ is assigned
length dimension 1, the time $t$ assigned the unspecified length 
dimension $\ep$, and the  disk amplitude $W$ the length dimension $-\eta$. 
We assume $x$, as the coupling constant conjugate to the 
length $\ell$ has length dimension ${-1}$. We leave the length 
dimension of $g_s$ unspecified as $-\del$.

Denote the dimensionless lattice variables and observables  
by $t_{reg}$, $W_{reg}$ etc.
We thus have  $t_{reg} = t/a^{\ep}$. Similarly
with $W$ and $W_{reg}$: $W_{reg} = W \, a^\eta$, except that we allow for  
the possibility that $W_{reg}$ is 
not scaling when $a \to 0$ (i.e. that $\eta > 0$). 
Although at first sight a little strange 
this was precisely what happened  in the ordinary scaling limit 
of the matrix models, as we noted in the discussion above (see eq.\ \rf{2.3}
which shows that in the ordinary scaling limit we have $\eta = 3/2$). 
Thus we will allow for this possibility. We can summarize as follows: 
\bea
W_{reg} &\xrightarrow[a\to 0]{}& a^{\eta}\, W_\lam(x),\hspace{17mm} 
\eta < 0, 
\label{5.5}\\
t_{reg} &\xrightarrow[a\to 0]{}&  t/a^\ep,\hspace{25mm}{\ep =1}.
\label{5.6}\\
&& \nonumber\\
&& \nonumber\\
W_{reg} &\xrightarrow[a\to 0]{}& {\rm const.} +a^{\eta}\, W_\lam(x), 
~~{\eta=3/2} \label{5.7}\\
t_{reg}& \xrightarrow[a\to 0]{}&  t/a^\ep, \hspace{24mm} {\ep=1/2},
\label{5.8}
\eea
\begin{figure}[t]
\centerline{\scalebox{0.5}{\rotatebox{0}{\includegraphics{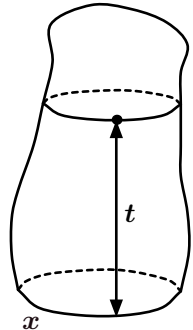}}}}
\caption{The differentiation of the disk amplitude with respect to the 
cosmological constant leads to the disk amplitude with one marked point.
Such a geometry has the unique decomposition shown in the figure, 
where the loop has a geodesic distance $t$ from the boundary loop.
The equation graphically represented on the figure is \rf{5.9}.}
\label{fig6}
\end{figure}
The values of the exponents $\ep$ and $\eta$ are written to the right in
equations \rf{5.5}-\rf{5.8}. They tell us that when $\eta < 0$, i.e.\ 
when $W_{\lam,g_s}(x)$ scales, then $\ep =1$, while if 
$\eta > 0$ then $\ep =1/2$ and $\eta = 3/2$. 
The reason these exponent are uniquely determined is that we have 
the  geometric picture shown in Fig.\ \ref{fig6}, which couples $W$ and $G$ 
and leads to consistency relations for the scaling of $W$:
\beq\label{5.9}
-\frac{\prt W_{\lam,g_s}(x)}{\prt \lam} =
 \int_0^\infty d t \int_{0}^{\infty} 
d\ell\ G_{\lam,g_s} (x,\ell;t)\, \ell W_{\lam,g_s}(\ell).
\eeq
We know the scaling dimension of $ G_{\lam,g_s} (x,\ell;t)$. It is zero.
It all goes back to the fact that $G$ as a propagator has to satisfy the 
composition rule:
\beq\label{5.10}
G(\ell_1,\ell_2;t_1+t_2) = \int_0^\infty d \ell \; 
G(\ell_1,\ell;t_1)G(\ell,\ell_2;t_2),
\eeq
valid for both $G_{\lam,g_s}(\ell_1,\ell_2;t)$ and 
$G_{\lam}^{(0)}(\ell_1,\ell_2;t)$. This means that 
$G(\ell_1,\ell_2;t)$ has to scale like $a^{-1}$ and the 
Laplace transform $G(x,\ell;t)$ thus as $a^0$. Combined with that fact
that the cosmological constant $\lam$ has length dimension  ${-2}$, 
we are led to right side values of $\ep$ and $\eta$ 
in eqs.\ \rf{5.5}-\rf{5.8}. It is 
a beautiful example of the constraints imposed by {\it quantum geometry},
and it is remarkable that one is able to derive the non-trivial, 
ordinary matrix model  value $\eta =3/2$ in eq.\ \rf{2.3} from such 
simple geometric considerations.

\begin{figure}[t]
\centerline{\scalebox{0.5}{\rotatebox{90}{\includegraphics{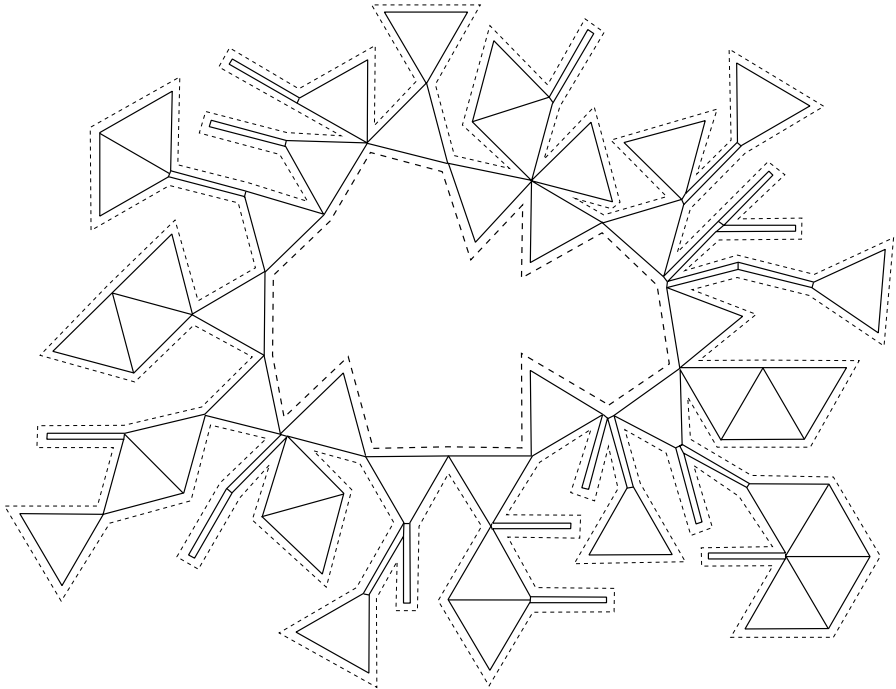}}}}
\caption{Part of a ``typical'' triangulation of the disk which one 
will encounter in the ``ordinary scaling'' limit of the matrix models, where
equations \rf{5.7} and \rf{5.8} are satisfied. The dashed and dotted lines
represent two ``spatial'' curves separated by one (proper)-time step, plus
all the baby universes which are cut off at the this time step.}
\label{fig7}
\end{figure}
If we choose the solution \rf{5.5}-\rf{5.6}
then we can solve the geometric equations \rf{5.2} and \rf{5.4}.
A glance on eq.\ \rf{5.4} shows that {\it if} the term involving 
$g_s$ is going to play a role we have to take $\del =3$, the 
same result as in the new scaling limit of the matrix models. If we 
make this choice the solution for  $W_{\lam,g_s}(x)$ is precisely the 
one given by the new scaling limit of the matrix models. Thus the 
new scaling limit has indeed a geometric representation in terms of 
random surfaces, and even a regularized 
lattice representation where a lattice spacing $a$ is taken to zero.
All this is discussed in detail in \cite{cdt1}-\cite{cdt2}. 
The main characteristic is
here that the geodesic distance (the proper time $t$) has canonical scaling
dimension, identical to that of space and there is a smooth limit $g_s \to 0$
where one obtains the original CDT solution \cite{al}.

If we choose the solution \rf{5.7}-\rf{5.8} and solve the geometric 
equations \rf{5.2} and \rf{5.4} we find a completely different solution.
It is characterized by a different scaling of the geodesic distance
or proper time $t$. We see that eq.\ \rf{5.4} is only  non-trivial 
{\it if} we choose the dimension of $g_s$ to be zero, and the 
first term on the rhs of eq.\ \rf{5.4} is then irrelevant. The 
equation is thus reflecting an excessive branching off of 
baby universes. There is nothing else!  Thus, looking
at Fig.\ \ref{fig4} and \ref{fig5}, we have no unshaded parts of the 
graphs. They simply play no role in the scaling limit where 
the lattice spacing $a \to 0$. The typical geometry which
appears in the path integral will be very fractal and will have 
Hausdorff dimension $d_h=4$, not $d_h=2$ as one would expect 
from a ``nice'' two-dimensional geometry. This is described in 
detail in \cite{hausdorff}. 
The wild branching of baby universes is illustrated in Fig.\ \ref{fig7}.
This limit with $d_h=4$ corresponds to the ``ordinary'' scaling limit
of the matrix model.

\begin{figure}[t]
\centerline{\scalebox{0.4}{\rotatebox{0}{\includegraphics{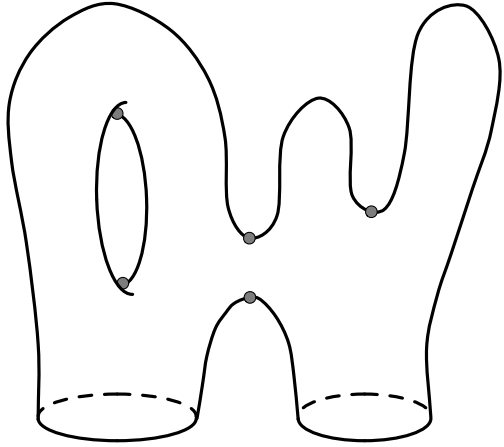}}}}
\caption{A ``typical'' geometry (in the continuum limit) of the disk which one 
will encounter in the ``new'' scaling limit of the matrix models, where
equations \rf{5.5} and \rf{5.6} are satisfied. We have show a surface where
the topology of the surface is also changed from that of a simple disk to 
a disk with a handle, see discussion in the main text.}
\label{fig8}
\end{figure}

This is in sharp contrast to the solution provided 
by conditions \rf{5.5}-\rf{5.6}. There the typical geometry has
$d_h=2$ and the total number of baby universes created is finite.
This implies that a typical geometry will look as shown in Fig.\ \ref{fig8}.
In Fig.\ \ref{fig8} we have drawn an amplitude which 
is more complicated than the disk amplitude (we have two entrance loops, 
and the surface also have a handle, i.e.\ it is a higher genus surface.
This gives us the opportunity t0 emphasize that while we have mainly 
discussed geometries of the simplest topology, the whole discussion 
of the new scaling generalizes to any genus surface and with 
any number of boundary loops, and it matches 
precisely the $1/N^2$ expansion in the  matrix models of expectation
values of multiple trace operators \cite{cdt1}-\cite{cdt3}:
\beq\label{5.11}
\om(z_1,\ldots,z_n)= \la \frac{1}{N}\tr \frac{1}{z_1-\phi} \cdots \frac{1}{N} 
\tr\frac{1}{z_n-\phi} \ra_{connected}. 
\eeq 
These multiple trace operators are obvious generalizations
of the single trace operator defined in eq.\ \rf{1.7}.    

\section*{Unfinished stuff}

For the ordinary matrix models we have a description 
of conformal matter coupled to 2d quantum gravity by 
multicritical one-matrix models and by two-matrix models.

Similarly it is easy to couple matter to the ``plain'' CDT model.
Ising models and multiple Pott models coupled to CDT 
have been studied numerically \cite{cdtmatter}.

Now that we have a matrix model description of the generalized CDT models
it seems natural to apply the same technique as was applied for 
the ordinary matrix models and in this way use matrix models 
to study the matter systems defined on the CDT-like set of random surfaces.
From the computer simulations referred to, it seems that one
obtain the flat space-time exponents. It would be 
very interesting if one could obtain a simple proof of 
the any conformal field theory exponent from a matrix integral.
It would provide us with an explicit realization 
of these critical systems, even at a regularized level. 
 Work in this direction is in progress.

\section*{Acknowledgment}
I thank   R. Loll, Y. Watabiki, W. Westra and S. Zohren for 
a wonderful collaboration in trying to understand how to 
take a new scaling limit of the matrix models. All
mistakes in this article are due to me. I also 
thank Utrecht University as well as the Perimeter Institute,
where part of this work was done, for hospitality and financial support.


\end{document}